\documentclass[pre,showpacs,superscriptaddress,twocolumn,longbibliography]{revtex4-1}

\usepackage{hyperref}
\usepackage{color}
\usepackage[usenames,dvipsnames]{xcolor}
\usepackage{amsmath,amsthm,amssymb}
\usepackage[normalem]{ulem}
\usepackage{graphicx}
\usepackage{float}
\usepackage{epsfig}
\usepackage{bm}
\usepackage{mathrsfs}
\usepackage{multirow}
\usepackage[all]{xy}
\usepackage{pbox}
\usepackage{verbatim}
\usepackage{braket}
\usepackage{mathtools}
\usepackage{bm}
\usepackage{tikz}
\usepackage{xcolor}
 
\usepackage{mathtools}

\usepackage{mathtools}
\DeclareMathOperator{\Tr}{Tr}

\newcommand{\er}[1]{Eq.~\eqref{#1}}

\newcommand{\era}[2]{Eqs.~(\ref{#1}) and (\ref{#2})}

\newcommand{\beq}{\begin{equation}}
\newcommand{\eeq}{\end{equation}}

\newcommand{\W}{\mathbb W}
\newcommand{\R}{\mathbb R}
\renewcommand{\P}{\mathbb P}

\begin{document}  

\title{Dynamics and large deviation transitions of the XOR-Fredrickson-Andersen kinetically constrained model}

\author{Luke Causer}
\affiliation{School of Physics and Astronomy, University of Nottingham, Nottingham, NG7 2RD, UK}
\affiliation{Centre for the Mathematics and Theoretical Physics of Quantum Non-Equilibrium Systems,
University of Nottingham, Nottingham, NG7 2RD, UK}
\author{Igor Lesanovsky}
\affiliation{Institut f\"ur Theoretische Physik, Universit\"at T\"ubingen, Auf der Morgenstelle 14, 72076 T\"ubingen, Germany}
\affiliation{School of Physics and Astronomy, University of Nottingham, Nottingham, NG7 2RD, UK}
\affiliation{Centre for the Mathematics and Theoretical Physics of Quantum Non-Equilibrium Systems,
University of Nottingham, Nottingham, NG7 2RD, UK}
\author{Mari Carmen Ba\~nuls}
\affiliation{Max-Planck-Institut f\"ur Quantenoptik, Hans-Kopfermann-Str.\ 1, D-85748 Garching, Germany}
\affiliation{Munich Center for Quantum Science and Technology (MCQST), Schellingstr.\ 4, D-80799 M\"unchen}
\author{Juan P. Garrahan}
\affiliation{School of Physics and Astronomy, University of Nottingham, Nottingham, NG7 2RD, UK}
\affiliation{Centre for the Mathematics and Theoretical Physics of Quantum Non-Equilibrium Systems,
University of Nottingham, Nottingham, NG7 2RD, UK}

\begin{abstract}
We study a one-dimensional classical stochastic kinetically constrained model (KCM) inspired by Rydberg atoms in their ``facilitated'' regime, where sites can flip only if a single of their nearest neighbours is excited. We call this model ``XOR-FA'' to distinguish it from the standard Fredrickson-Andersen (FA) model. 
We describe the dynamics of the XOR-FA model, including its relation to simple exclusion processes in its domain wall representation. The interesting relaxation dynamics of the XOR-FA
is related to the prominence of large dynamical fluctuations that lead to phase transitions between active and inactive dynamical phases as in other KCMs. By means of numerical tensor network methods we study in detail such transitions in the dynamical large deviation regime. 
\end{abstract}

\maketitle

\section{Introduction}

Systems with constraints often display interesting cooperative dynamics \cite{Fredrickson1984, Palmer1984, Jackle1991, Kob1993}. This is true both in classical and quantum settings. Broadly speaking there are three classes of constrained systems. One is that of problems where state space is constrained. The canonical example is lattice coverings, for example dimers on a square lattice \cite{Rokhsar1988,Alet2005, Alet2006, Syljuasen2006, Henley2010, Oakes2016, Oakes2018}. In such systems, the constrained nature of configuration space implies constraints in the allowed transitions, making both their classical and quantum dynamics very rich. A second class encompasses systems where constraints in the dynamics are emergent, such as in classical and quantum ``fracton'' models where the motion of certain effective excitations is severely impeded \cite{Vijay2015, Vijay2016, Prem2017, Song2019, Nandkishore2019}. A third class comprises systems known as kinetically constrained models (KCMs) with explicit constraints in the allowed dynamical transitions. Here we focus on KCMs. 

KCMs were first introduced  \cite{Fredrickson1984,Palmer1984} in the 1980s as models of classical glasses. The ones studied most throroughly, such as the Fredrickson-Andersen (FA) \cite{Fredrickson1984} and East models \cite{Jackle1991}, are stochastic lattice spin systems with the interesting combination of a trivial thermodynamics and a strongly fluctuating cooperative dynamics (under appropriate conditions - typically low temperatures and/or high densities) due to the constraints. 
For reviews on classical KCMs see e.g.~\cite{Ritort2003,Garrahan2011,Garrahan2018}. Like their classical counterparts, quantum KCMs also display complex non-equilibrium dynamics, both under closed unitary \cite{Horssen2015,Lan2018,Pancotti2020} or open dissipative \cite{Olmos2012} evolution. 

Modelling dynamics via KCMs can be motivated in many different areas. For example, in classical soft matter, specifically for glasses \cite{Chandler2010,Biroli2013}, kinetic constraints are meant to encode the local steric interactions of dense fluids. Another application is in the context of ensembles of Rydberg atoms in optical lattices, modelled as a collection of local two-level systems (representing for each atom their ground and some high-lying Rydberg state). When driven on resonance, due to ``Rydberg blockade'' \cite{Browaeys2019}, their dynamics is subject to a kinetic constraint where an atom can change state only if all their nearest neighbours are in their ground state. In a one-dimensional lattice such constraint gives rise to the much studied PXP model \cite{Fendley2004,Lesanovsky2011,Bernien2017,Turner2018}, the quantum counterpart of the classical ``two-spin facilitated'' FA model \cite{Ritort2003}. 

Here we study a one-dimensional classical KCM which to our knowledge has not been considered in the past. We call it the XOR-FA model to distinguish it from the standard FA model (i.e.\ the ``one-spin facilitated'' FA model). The kinetic constraint in the XOR-FA is such that a spin can flip only if one of its nearest neighbours is in the excited state, but not if both are (which is allowed in the FA). Such condition makes the XOR-FA more constrained than the standard FA model. Conversely, the XOR-FA is less constrained than the PXP, whose transitions require the two neighbouring sites to be simultaneously down. The constraint in the XOR-FA model can be motivated by Rydberg atoms in their ``facilitated'' (or ``anti-blockade'') regime \cite{Ates2007,Amthor2010,lesanovsky2013, Lesanovsky2014, Hoening2014, Urvoy2015, Valado2016, Marcuzzi2017, Ostmann2019, Wintermantel2020, Mazza2020}: when driven out of resonance, specifically when blue-detuned, conditions can be such that an atom may change state only if a single neighbour is in the excited state, but not both. 

The paper is organised as follows. In Sec.\ II we introduce the XOR-FA model and discuss its connection to simple exclusion processes. In Sec.\ III we consider the relaxation dynamics of the model. In Sec.\ IV we study the dynamical large deviations by means of numerical tensor networks, and show the existence of a phase transition between active and inactive dynamical phases. In Sec.\ V we draw comparisons between the FA, XOR-FA and PXP models. In Sec.\ VI we give our conclusions.

\section{Model} 

We consider a system of binary variables $n_j = 0,1$ (we call these states down/up or unexcited/excited) on the sites $i=1,\ldots,N$ of a one-dimensional lattice (with boundary conditions to be specified below). Similarly to other KCMs \cite{Ritort2003,Garrahan2010} the dynamics will be that of singe-spin flips subject to a constraint. Specifically, the allowed transitions are
\beq
\begin{array}{rcl}
001 \to 011 & & \text{rate}=c \\ 
011 \to 001 & & \text{rate}=1-c \\ 
100 \to 110 & & \text{rate}=c \\ 
110 \to 100 & & \text{rate}=1-c, \\ 
\end{array} 
\label{rules}
\eeq
where $c\in(0, 1)$.
That is, a site can flip only if both nearest neighbouring sites are in different states. This means that the constraint is a boolean XOR operation on the nearest neighbours of the site that is attempting to flip. 
We therefore call this model the XOR-FA (short for XOR-Fredrickson-Andersen) to distinguish it from the standard Fredrickson-Andersen (FA) model, where a site can flip if either of its nearest neighbours is up, which in this nomenclature would correspond to the OR-FA (while the PXP would be the AND-FA).

The generator of the continuous-time Markov dynamics is the operator
\beq
\W = \sum_{j=1}^N \P_j \left[ c \sigma_j^+ + (1-c) \sigma_j^- - c(1-n_j) - (1-c) n_j \right] \, ,
\label{W}
\eeq
where $\sigma_j^\pm$ are Pauli operators acting on site $j$, $n_j = \sigma_j^+ \sigma_j^-$, and the kinetic constraint $\P_j$ on site $j$ reads, 
\beq
\P_j = \left( n_{j-1} + n_{j+1} - 2 n_{j-1} n_{j+1} \right) =
\frac{1}{2} \left( 1 - \sigma_{j-1}^z \sigma_{j+1}^z \right) \, ,
\label{cons} 
\eeq
where $\sigma_j^z = 2 n_j - 1$. The operator \er{cons} enforces the impossibility of the transitions ruled out in \er{rules}. Note that \er{W} has an explicit symmetry between up/down spins and is unchanged under the transformation $c \to 1-c$ and $n_{j} \to 1 - n_{j}$.

Dynamics with the kinetic constraint \er{cons} is naturally motivated \cite{Ostmann2019} in quantum many-body systems, specifically in the context of Rydberg atoms in their facilitated/anti-blockade regime \cite{Ates2007,Amthor2010,lesanovsky2013, Lesanovsky2014, Hoening2014, Urvoy2015, Valado2016, Marcuzzi2017, Ostmann2019, Wintermantel2020, Mazza2020}, whereby an up (down) spin represents an atom in its excited Rydberg (ground) state, and the drive is such that an atom can get excited resonantly only when one of its nearest neighbours is also excited, but not both (as that would make the transition off resonant). The constraint \er{cons} has also been studied in certain quantum spin chains \cite{Suzuki1971,Borla2020} in particular in relation to ``quantum scars'' (special non-thermal states in constrained quantum systems \cite{Shiraishi2017,Moudgalya2018,Turner2018}) \cite{Iadecola2020,Yang2020,Mark2020}, and additionally in the context of quantum cellular automata \cite{Gopalakrishnan2018c}. Our aim here is to consider the classical stochastic dynamics of a system with such a constraint, thus extending the set of known KCMs.

\subsection{Conservation of the number of domain walls and relation to simple exclusion processes}

The dynamical rules \er{rules} impose a conservation law in the dynamics, that of the total number of domain walls (DWs) \cite{Ostmann2019,Suzuki1971}. Consider two neighbouring domains of, say, up and down spins
\[
\cdots 11110000 \cdots
\]
Due to the constraint \er{cons} the only allowed changes are to the spins next to the DW, since inside the domains both neighbours to every spin are the same. This means that the possible moves are
\[
\begin{array}{lcr}
& & \cdots 111{\underline 0}0000 \cdots \\
\cdots 11110000 \cdots & 
\begin{array}{c}
\nearrow \\
\searrow \\
\end{array}
& \\
& & \cdots 1111{\underline 1}000 \cdots \\
\end{array}
\]
where we have underlined the sites that changed in each allowed transition. 

We can perform a duality transformation to have an explicit DW representation of the problem. We write
\begin{align}
\sigma_j^x = \, & 
X_{j} X_{j+1}
\label{ctx} ,
\\
\sigma_j^y = \, & 
(-1)^{j+1}\prod_{k=1}^{j-1} Z_{k}Y_{j}X_{j+1}
\label{cty} ,
\\
\sigma_j^z =\, & (-1)^{j+1} \prod_{k=1}^j Z_k
\label{ctz} ,
\end{align}
where $X_j,Y_j,Z_j$ are Pauli operators for the DW between sites $j-1$ and $j$. Notice that this is a canonical (rather than unitary) transformation that preserves the commutation relations between the Pauli operators. The generator in this representation is then
\begin{align}
    \W^{\rm DW} = \sum_{j=1}^N & \ \P_j^{\rm DW} 
     \bigg [ 
    \frac{1}{2} X_j X_{j+1} 
    \nonumber
    \\
    &
    - i \left(\frac{1}{2}-c\right)(-1)^{j+1}\prod_{k=1}^{j-1}Z_{k}Y_{j}X_{j+1}
    \phantom{\prod_{k=1}^j}
    \label{WXX}
    \\
    & 
    - \left(\frac{1}{2}-c \right) (-1)^{j+1} \prod_{k=1}^j Z_k - \frac{1}{2}
    \bigg]
    \nonumber ,
\end{align}
where the constraint is
\beq
\P_j^{\rm DW} = \frac{1}{2} \left( 1 - Z_j Z_{j+1} \right) 
\eeq
and we have used the superscript ``DW'' to indicate operators in the domain wall representation. Combining the factors we can simplify the generator to 
\begin{align}
    \W^{\rm DW} = & \ \sum_{j=1}^N \frac{1}{2}
     \bigg [ 
    \frac{1}{2}\left(X_j X_{j+1} + Y_{j}Y_{j+1}\right)
    \nonumber
    \\
    &
    - i \left(\frac{1}{2}-c\right)(-1)^{j+1}\prod_{k=1}^{j-1}Z_{k}\left( Y_{j}X_{j+1} - X_{j}Y_{j+1} \right)
    \nonumber
    \\
    & 
    + \frac{1}{2}\left( Z_{j}Z_{j+1} - 1\right)
    \label{WDW}
    \\
    &
    - \left(\frac{1}{2}-c \right) (-1)^{j+1} \prod_{k=1}^{j-1} Z_k \left(Z_{j} - Z_{j+1} \right) 
    \bigg]
    \nonumber .
\end{align}
The conservation law is now explicit, as the kinetic term simply corresponds to DW hopping. That is, we conserve the quantity $\mathbb{N}_{\rm DW}= \frac{1}{2}\sum_{j} \mathbb{I} + Z_j$. For the special case of $c=1/2$, the generator simplifies to that of the symmetric simple exclusion process (SEP), \cite{Blythe2007, Mallick2015}
\beq
\W^{{\rm DW}}_{c=1/2} = \sum_{j=1}^N \frac{1}{4} 
\left( X_j X_{j+1} + Y_j Y_{j+1} + Z_j Z_{j+1} - 1 \right).
\label{WSEP}
\eeq
In the XOR-FA language this is the ``infinite temperature limit'', where the cost of creating and destroying an excitation is the same. Away from $c=1/2$, the dependence of $c$ in the DW-representation generator \er{WDW} encodes the fact that moving left or right a DW depends on whether it is energetically favourable to extend or contract the corresponding spin domain. 

For $c\neq 1/2$ the generator \er{WDW} corresponds to a SEP with particles with alternating asymmetries in their hopping rates. That is, we have a model where particles (DWs) can hop to neighbouring sites if the sites are not already occupied: the odd particles (DWs) hop left with rate $c$ and right with rate $1-c$, while the even particles hop left with rate $1-c$ and right with rate $c$. Since particles (DWs) cannot cross due to the exclusion, these rates are maintained. This is a special case of the general model introduced in Ref.~\cite{Evans1996}, where each particle is given an individual hopping rate which is maintained under the dynamics. Independently from Ref.~\cite{Evans1996}, this exclusion process was studied in Ref.~\cite{Sidoravicius1998}. In that paper the authors use a transformation onto a representation which coincides to our spin model, allowing them to find a hydrodynamic limit with a non-trivial diffusion rate for the exclusion process with alternating hopping rates.

\section{Equilibrium and relaxation}

\subsection{Equilibrium properties}

We consider the XOR-FA with $N$ sites, $N_{\rm DW}$ domain walls and periodic boundary conditions (PBC), which formally corresponds to setting $n_{0} = n_{N}$. The dynamics generated by \er{W} obeys detailed balance and therefore any initial condition eventually relaxes to an equilibrium state. Since the dynamics conserves the number of DWs, there is one such equilibrium probability for each DW sector. For PBC the number of DWs is even, and the sectors can be classified by the number $p$ of up/down (one/zero) domains, $p = N_{\rm DW}/2$. One can then construct the equilibrium state within each sector in the following way.

\begin{figure}[t]
    \centering
    \includegraphics[width=\linewidth]{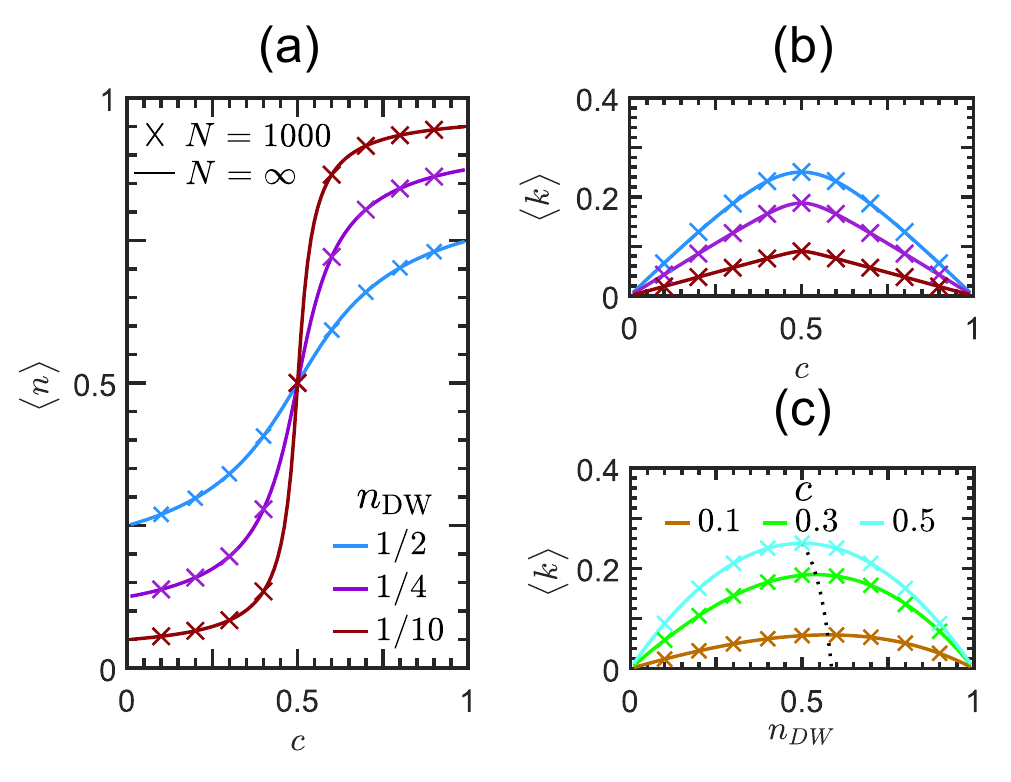}
    \caption{\textbf{Equilibrium properties of the XOR-FA model.} We show various properties of the XOR-FA at equilibrium for both finite size systems ($N = 1000$, symbols) obtained via Monte Carlo simulations, and $N = \infty$ (solid lines) obtained through the analytical considerations from the appendix. (a) The average excitation density for various DW fillings as a function of $c$. (b) The average dynamical activity $\braket{k}$ as a function of $c$ for the same DW fillings of (a). (c) The average dynamical activity $\braket{k}$ as a function of DW filling $n_{\rm DW}$ for various $c$. Note that the peak in the dynamical activity with respect to $n_{\rm DW}$ varies with $c$ (dotted line).}
    \label{fig:relaxation1}
\end{figure}

Consider a configuration for fixed $p$ where the zero (or down) domains and the one (or up) domains have lengths $d_{m}$ and $u_{m}$ respectively for $m = 1, \dots, p$, with the first domain being a down one,
\[
    \ket{0 .. 0_{d_{1}} 1 .. 1_{u_{1}} \dots 0 .. 0_{d_{p}} 1 .. 1_{u_{p}}} \, .
\]
Note that the total length of the domains must be equal to the system size, so in the state above we have 
\begin{equation}
\sum_{m=1}^{p} ( d_{m} + u_{m}) = N \,
\label{sum1}
\end{equation}
and each domain must have at least one site, so that 

\begin{equation}
d_{m}, u_{m} \geq 1 \;\; \forall m \, .
\label{sum2}
\end{equation}

We now define a state which is the translationally invariant superposition of all possible translations of the state above,
\[
    \ket{d_{1}, \, u_{1}, \, \dots , \, d_{p},\, u_{p}} = 
    \sum_{m=1}^{N} {\mathbb T}^{m}   
    \ket{0 .. 0_{d_{1}} 1 .. 1_{u_{1}} \dots 0 .. 0_{d_{p}} 1 .. 1_{u_{p}}} \, ,
\]
where the operator ${\mathbb T}$ shifts the chain by a single site.

The equilibrium probability vector for the sector with $2p$ DWs is given by
\beq
\begin{split}
\ket{{\rm eq}_p} = {\cal N}
\sum_{d_{1}=1}^\gamma \cdots
\sum_{d_{p}=1}^\gamma 
\sum_{u_{1}=1}^\gamma \cdots
\sum_{u_{p}=1}^\gamma 
\delta \left( d_{1} + \cdots + u_{p} - N \right) \\
(1-c)^{\sum_m d_{m}} c^{\sum_m u_{m}} 
\ket{d_{1}, \, u_{1}, \, \dots , \, d_{p},\, u_{p}}
\end{split}
\label{SS}
\eeq
where $\gamma = N - 2p + 1$ and ${\cal N}$ is a normalization constant. One can check that the state \er{SS} is annihilated by all terms of the generator \er{W}. This state corresponds to the equilibrium state with non-interacting energy $E = \sum_j n_j$ (i.e., each up spin costs a unit of energy) at temperature $T$ such that $c=e^{-1/T}/(1+e^{-1/T})$, and subject to the conditions \era{sum1}{sum2}. 

We now study the basic properties of the equilibrium state \er{SS}. In Fig.~\ref{fig:relaxation1} we show two average observables in equilibrium. The first one is the average excitation density, $\braket{n} = N^{-1}\sum_{i}\braket{-| n_{i} | {\rm eq}_{p}}$,
where $\bra{-} = \sum_{\bm{n}}\bra{\bm{n}}$ is the flat state and $\bra{\bm{n}} = \bra{n_{1}, \dots, n_{N}}$, see 
Fig.~\ref{fig:relaxation1}(a).  We show $\braket{n}$ for several values of the filling fraction defined as $n_{\rm DW}=2p/N$
(note that the mean domain length is $1 /n_{\rm DW}$).
The symbols are numerical results from standard Monte Carlo simulations. Note that in contrast to the FA or East models \cite{Ritort2003,Garrahan2010}, $\braket{n}$ is not just equal to $c$, due to the conservation of the number of DWs. Figure~\ref{fig:relaxation1}(a) shows the agreement of the numerics with an analytical prediction in the $N\to\infty$ limit described in the Appendix.

The second observable coincides with the average dynamical activity (per site) in equilibrium, $\braket{k}$. While the  activity is an observable at the level of trajectories (see Sect.~IV below for further details), its average in equilibrium is given by the average escape rate per site, which is a static observable \cite{Garrahan2009}. The escape rate operator $\R$ is (minus) the diagonal part of the generator \er{W}.
Since $\R$ is a local operator we can also obtain analytically its equilibrium average in the large size limit, see Appendix. In Fig.~\ref{fig:relaxation1}(b) we show the agreement between $\braket{k}$ from simulations and the analytic result. Note that $\braket{n}$ and $\braket{k}$ are symmetric around $c=1/2$ as functions of $c$ as a consequence of the up/down symmetry of the model, cf.\ \er{W} (while there is no corresponding symmetry in terms of DW filling $n_{\rm DW}$).

\begin{figure}[t]
    \centering
    \includegraphics[width=\linewidth]{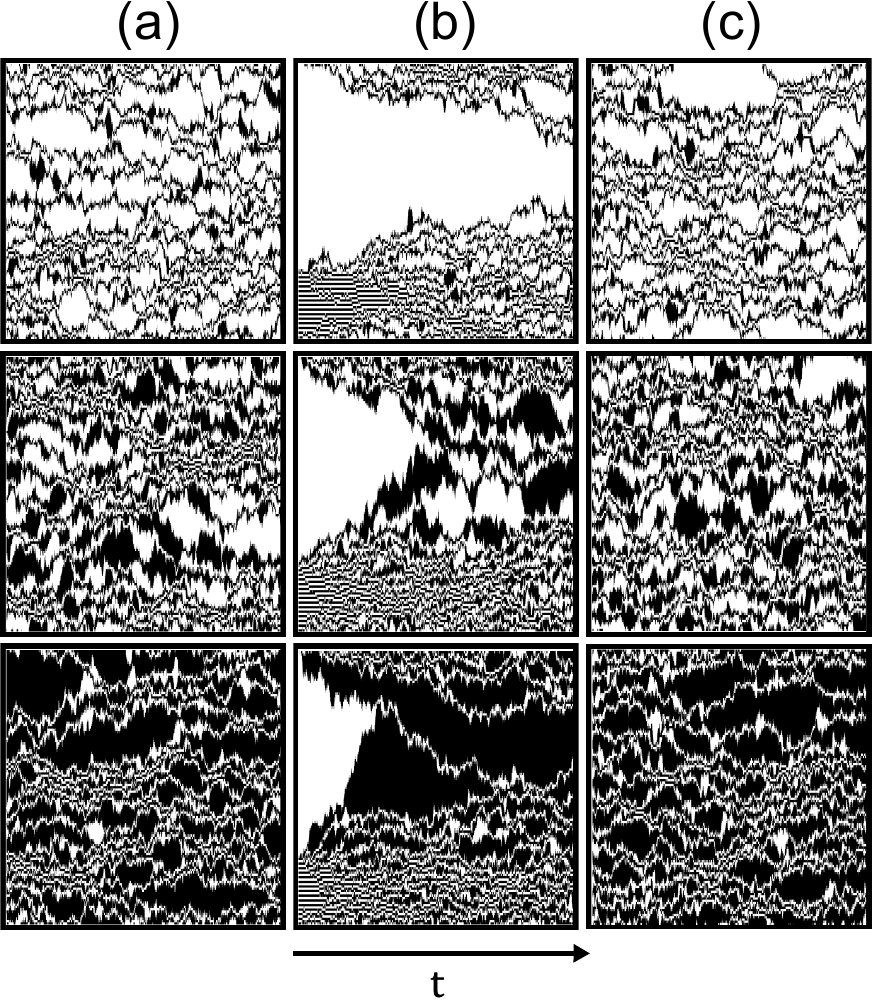}
    \caption{\textbf{Trajectories of the XOR-FA model.} Representative trajectories from continuous-time Monte Carlo simulations using generator \er{W} with a time of $t = 10^{3}$. The rows are for different values of $c$, with $c = 0.4$ (top), $c = 0.5$ (middle), $c = 0.6$ (bottom). All trajectories are at quarter filling, $n_{\rm DW} = 1 / 4$. 
    The columns correspond to different initial conditions: (a) typical equilibrium configuration, (b) DWs maximally clustered, (c) DWs maximally spread out.  Column (b) shows the slowest approach to equilibrium. 
    }
    \label{fig:relaxtation2}
\end{figure}

\subsection{Relaxation dynamics} 

The dynamics of the XOR-FA model is entirely determined by expansion and contraction of the domains (or the movement of domain walls which cannot cross). The system behaves like an ``accordion''. Depending on the value of $c$ there may be an energetic preference to expand or contract domains of one orientation or the other. Fig.~\ref{fig:relaxtation2} shows typical trajectories sampled from the XOR-FA model by running continuous time Monte Carlo at various conditions. The trajectories of the figure are at a quarter filling for three values of $c$. The top row of Fig.~\ref{fig:relaxtation2} is for $c=0.4$, the middle one for $c=0.5$, and the bottom one, $c=0.6$. The columns correspond to different initial conditions. Column (a) shows equilibrium trajectories, i.e., those that start from an initial condition sampled from \er{SS}. They show pronounced space-time fluctuations in the dynamics associated to the breathing of domains. Column (b) corresponds to the most unfavourable initial state, where DWs are maximally clustered. Relaxation to equilibrium in this case is slow, as DWs within the bulk of the cluster cannot move until the DWs on the outside of cluster diffuse away. Column (c) shows an opposite non-equilibrium initial condition, where DWs are maximally spread out. In this case relaxation to equilibrium is faster, cf.\ Fig.~\ref{fig:relaxtation2}(c). The large space-time fluctuations that are evident in these example trajectories anticipate the large deviation phase transitions that we uncover in the next section.

The different timescales involved in the relaxation of the XOR-FA model can be quantified using time-correlation functions. In particular we focus on two different correlators in the equilibrium dynamics. The first one is the auto-correlation function, $C(t)$, which measures how many sites that were in the excited state at time $0$ are also in an excited state at a later time $t$. Subtracting the disconnected part, and normalising so that it takes values between 1 and 0, it reads,
\beq
C(t) = \frac{1}{N} \sum_{j=1}^{N} \frac{\braket{n_{j}(t) n_{j}(0)} - \braket{n}^{2}} {\braket{n}-\braket{n}^{2}} \, ,
\label{AC}
\eeq
where the average is over realisations of the dynamics in equilibrium, i.e., starting from a configuration sampled from the equilibrium state \er{SS} and evolved according to \er{W}. 

The second correlator we study is the persistence function, $P(t)$, which quantifies the average probability for a randomly selected site to have not changed state up to time $t$. We can define it in terms of a local dynamical variable $p_j(t)$ at each site $j$, where $p_{j}(t) = 1$ if the site has never changed from its initial state at time $t$, and $p_{j}(t) = 0$ as soon as it changes for the first time. The resulting aggregate function is then 
\beq
P(t) =  \frac{1}{N} \sum_{j=1}^{N} \braket{p_j(t)} .
\label{pers}
\eeq
This function is automatically normalised between 1 at the initial time and 0 eventually when all sites flip at least once. 

\begin{figure}[t]
    \centering
    \includegraphics[width=\linewidth]{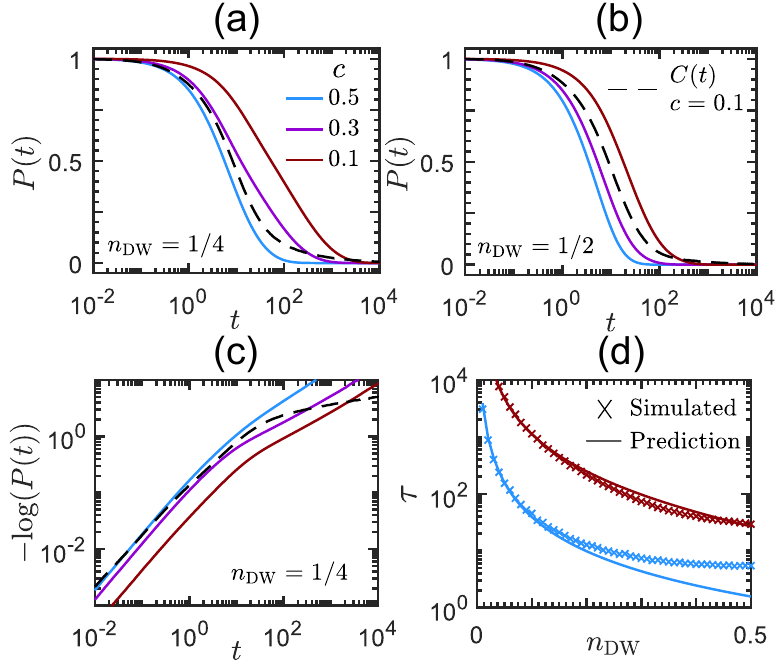}
    \caption{\textbf{Time correlations of the XOR-FA model.} All results are done for $N = 400$. We show the persistence function $P(t)$ plotted for (a) $n_{\rm DW} = 1/4$ and (b) $n_{\rm DW} = 1/2$. In both cases we show for various $c$. The black dashed line also shows the auto-correlator $C(t)$ for $c = 0.1$ to compare. 
    (c) We plot the same functions as shown for (a) but with a double-logarithmic scale on the ordinate.
    (d) We show the time taken $\tau$ for the persistence function to drop to $P(\tau) = e^{-1}$ (crosses) for $c=0.5$ (blue, bottom), $c=0.1$ (red, top) and various $n_{\rm DW}$. We also show our estimate $\tau$ (solid lines) given in \er{timescale}.
    }
    
    \label{fig:relaxtation3}
\end{figure}

In Fig.~\ref{fig:relaxtation3} we show results for time-correlators. We focus mostly on the persistence function as it better captures overall relaxation. Figure~\ref{fig:relaxtation3} shows $P(t)$ for various $c$ and two filling fractions of DWs, $n_{\rm DW} = 1/4$ (a) and $n_{\rm DW} = 1/2$ (b). 
For comparison we also show the auto-correlator for $c = 0.1$ (dashed). We see that decreasing $c$ away from $c = 0.5$ leads to slower relaxation times. The same can be said for decreasing the density of the DWs. 
Figure~\ref{fig:relaxtation3}(c) shows the same functions as in (a) but in a double-logarithmic scale on the ordinate. The change in slope in this representation emphasises the change from exponential decay at short times, to stretched exponential decay at long times
\footnote{A stretched exponential has the form $P(t) \sim \exp\left[-(t/t_{0})^{b}\right]$. For $c=0.1$ and $n_{\text{DW}} = 1/4$, we estimate the ``stretching'' parameter to be $b = 0.488$.}.

From the persistence function we can extract a characteristic relaxation time, $\tau$, customarily from the time the function decays to $e^{-1}$, that is, $P(\tau) = e^{-1}$. These times are shown in 
Fig.~\ref{fig:relaxtation3}(d) for two values of $c$ and as a function of the DW filling. Their behaviour can be understood approximately with simple heuristic arguments. 

We first note that for $c \ll 1$, we can treat the dynamics of the XOR-FA model as small up domains diffusing around a ``vacuum'' of down domains. To move, the up domain must first expand by exciting a neighbouring spin. This happens slowly at rate $c$. Following this, the domain then shrinks at rate $1-c \approx 1$. It can either shrink back to its original position, or shrink such that it shifts by one site across, each happening with equal probability. Thus we say it diffuses around the lattice with diffusion constant $D_{c} \approx c / 2$. 
The time taken for the system to fully relax can then be estimated as the time it takes for the DWs to diffuse from their original positions around the available space surrounding them, until they hit another DW. On average, the length of each zero domain is given by the average number of down spins split among the number of zero domains. Namely,
\beq
l_{0} = \frac{2}{n_{\rm DW}} (1 - \braket{n}) .
\eeq
It then follows that the timescale for the system to relax goes as
\beq
\tau \sim \frac{({l_{0} / 2})^{2}}{D_{c}} = \frac{2}{c} \frac{(1-\braket{n})^{2}} {{n_{\rm DW}}^{2}}
\label{timescale}
\eeq
for $c$ small. As Fig.~\ref{fig:relaxtation3}(d) shows, this prediction works well for $c$ small in the whole $n_{\rm DW}$ range, while for $c\approx0.5$ it qualitatively accounts for $\tau$ for small DW density \footnote{The diffusion constant used in \er{timescale} is only true for small $c$. For $c=0.5$, the diffusion constant can be estimated to be $D_{c} \approx c$. In this way, the estimate \er{timescale} only accounts for the results qualitatively where we have to re-scale by some constant.}.

Thinking of the dynamics in this way can also explain the two timescales in Fig.~\ref{fig:relaxtation3}(c). At some small time after $t\gtrsim0$, the first successful shift of domain(s) will occur. When this happens for $c\ll 1/2$, the original site is no longer excited, but the site next to it is. In the language of the persistence, this means two sites have flipped from their initial state. For both the persistence and the auto-correlator, this gives a fast initial relaxation, and as these are random uncorrelated events, the initial decay is exponential. Further successive moves of the domain only change at most one more site from its initial state (or in the case of the auto-correlator, will only slightly reduce the probability that the domain may end up in its original position). Thus the rate at which relaxation occurs is reduced, the time is longer, and the decay of the correlators is stretched as the relaxation becomes more collective.

\section{Dynamical Large Deviations and Matrix Product States}

In this section we study the statistics of trajectories of the XOR-FA model in the long-time regime where we can apply large deviation (LD) methods \cite{Touchette2009,Garrahan2018,Jack2020}. Recent work \cite{Banuls2019,Helms2019,Helms2020} has shown the effectiveness of numerical tensor network methods for studying the LDs of KCMs. Here, by means of numerical matrix product states (MPS) we are able to study the LDs of the XOR-FA for large systems to high accuracy. As we show below, the XOR-FA has a trajectory-space phase transition between between dynamical phases with very distinct characteristics, similar to what occurs in several other KCMs. 

\subsection{LDs and tilted generators}
The dynamical activity \cite{Lecomte2007,Garrahan2007,Garrahan2018,Maes2020} is a trajectory observable which counts the number of configuration changes (in our case the number of spin flips) in some given time. It is the natural trajectory observable to quantify the amount of motion in the dynamics. 
A question one can ask is what is the probability of observing the activity $K$ for trajectories $\omega_{t}$ which run for a total time $t$. The probability distribution for $K$ is given by 
\beq
P_{t}(K) = \sum_{\omega_{t}}\pi(\omega_{t})\, \delta[K(\omega_{t})-K] ,
\eeq
where $\pi(\omega_{t})$ is the probability of observing trajectory $\omega_{t}$. For long times this obeys a large deviation (LD) principle $P_{t}(K)\approx e^{-t\varphi(K/t)}$ where $\varphi(K/t)$ is the LD rate function \cite{Touchette2009}. One can also consider the moment generating function
\begin{equation}
\label{Zs}
    Z_{t}(s) = \sum_{K}P_{t}(K) \, e^{-sK} = \sum_{\omega_{t}}\pi(\omega_{t})\, e^{-sK(\omega_{t})} ,
\end{equation}
which also obeys a LD principle, $Z_{t}(s) \approx e^{t\theta(s)}$  where $\theta(s)$ is the scaled cumulant generating function (SCGF) whose derivatives at $s=0$ give the cumulants of $K$, scaled by time \cite{Touchette2009}. The SCGF plays the role of the thermodynamical free energy and is related to the LD rate function by a Legendre transform $\theta(s) = - \min_{k}[sk + \varphi(k)]$ \cite{Touchette2009}.

We can deform the generator given in Eq. (\ref{W}) by multiplying the off-diagonals by a factor of $e^{-s}$ to give the tilted generator,
\begin{align}
\label{Ws}
\W_{s} = \sum_{j=1}^N \P_j [ & e^{-s} \big(c \sigma_j^+ + (1-c) \sigma_j^-\big)  \\
& - c(1-n_j) - (1-c) n_j ] ,
\nonumber
\end{align}
whose largest eigenvalue is the SCGF $\theta(s)$ \cite{Touchette2009}. It has the associated left and right eigenvectors, $\bra{l_{s}}\mathbb{W}_{s} = \theta(s)\bra{l_{s}}$ and  $\mathbb{W}_{s}\ket{r_{s}} = \theta(s)\ket{r_{s}}$ respectively. As the dynamics obeys detailed balance, we can transform the generator into a Hermitian one by using a similarity transformation independent of $s$ \cite{Garrahan2009}. We first define the diagonal matrix $Q$ with matrix elements $\braket{\textbf{n}|Q|\textbf{n}} = (1 - c)^{N/2}[c / (1-c)]^{\sum_{i}n_{i}/2}$. The tilted Hamiltonian $\mathbb{H}_{s} = -Q^{-1}\mathbb{W}_{s}Q$ is then given by 
\beq
 \mathbb{H}_{s} = -\sum_{j=1}^N \P_j \left[ e^{-s}\sqrt{c(1-c)}\sigma_{j}^{x} - c(1-n_j) - (1-c) n_j \right] \, ,
\label{Hs}
\eeq
which has the ground state $\mathbb{H}_{s}\ket{\psi_{s}} = -\theta(s)\ket{\psi_{s}}$. As was done for the generator, one can write the tilted Hamiltonian in the DW representation
\begin{align}
    \mathbb{H}_{s}^{\rm DW} = -\sum_{j=1}^{N} \frac{1}{2} \bigg [ & e^{-s}\sqrt{c(1-c)}(X_{j}X_{j+1} + Y_{j}Y_{j+1})
    \nonumber
    \\
    &
     + \frac{1}{2}(Z_{j}Z_{j+1} - 1) 
     \label{HDW}
    \\
    &
    - \left(  \frac{1}{2} - c \right)(-1)^{j+1}\prod_{k=1}^{j-1}Z_{k}(Z_{j} - Z_{j+1}) \bigg]
    \nonumber .
\end{align}
The eigenvector $\ket{\psi_{s}}$ of $\mathbb{H}_{s}$ is related to the left and right eigenvectors of $\mathbb{W}_{s}$ by 
\beq
    \ket{\psi_{s}} = \sum_{\textbf{n}} \sqrt{l_{s}(\textbf{n})r_{s}(\textbf{n}}) \ket{\textbf{n}} ,
\eeq
where $l_{s}(\textbf{n}) = \braket{l_{s} | \textbf{n}}$ and $r_{s}(\textbf{n}) = \braket{\textbf{n} | r_{s}}$. Thus studying the LDs reduces to diagonilising Eq. (\ref{Hs}) to find $\theta(s)$ and $\ket{\psi_{s}}$.

\begin{figure*}[t]
    \centering
    \includegraphics[width=\linewidth]{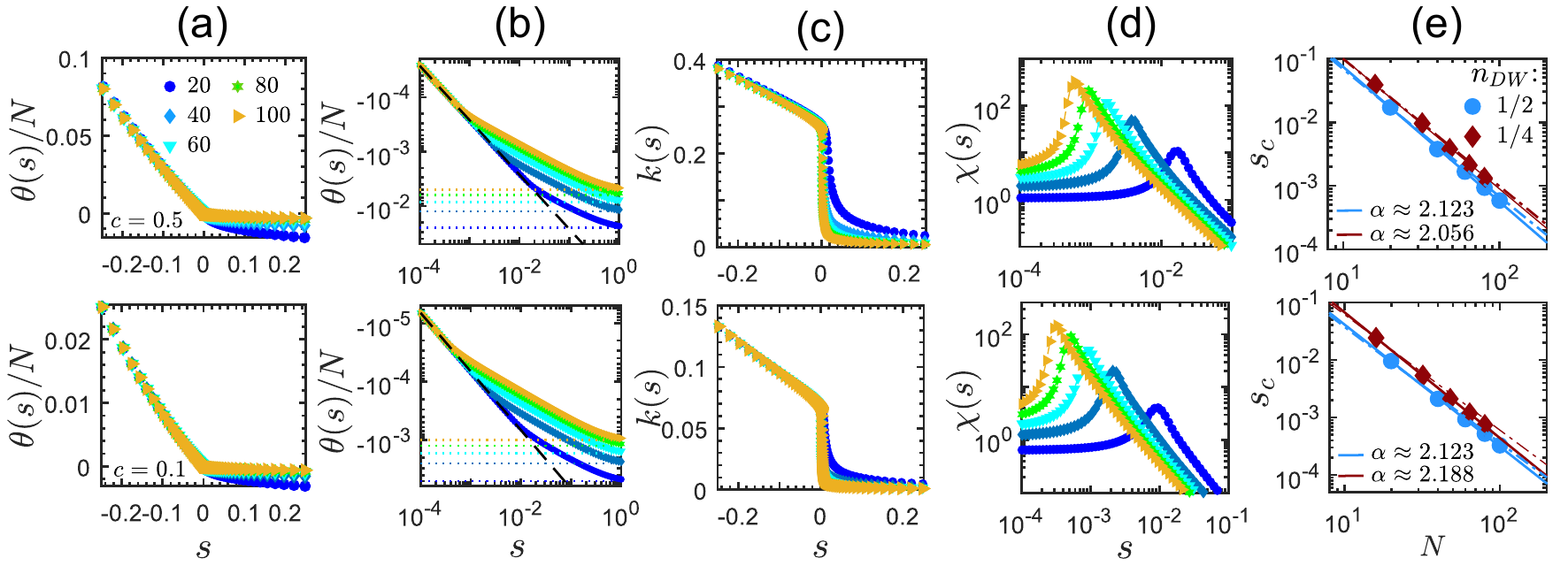}
    \caption{\textbf{First order phase transition in the SCGF.} We consider the finite size scaling of the phase transition for $N \in [20, 100]$ and $n_{\rm DW} = 1 / 2$. (a) A linear-linear plot of the SCGF $\theta(s) / N$. There is an apparent change in behaviour at some critical point $s_{c}(N) > 0$. (b) The SCGF $\theta(s) / N$ for $s>0$ on a log-log plot. On the first branch, $\theta(s) / N$ follows a LR (dashed line) up to $s_{c}(N)$ after which it follows another branch, which is no longer linear or scales with system size. The dotted lines show the value of the SCGF at $s\to\infty$. (c) The activity $k(s) = -\theta'(s) / N$ has a drop around $s = s_{c}(N)$ which becomes sharper with system size. (d) The dynamical susceptibility $\chi(s) = \theta''(s)$ which peaks at $s_{c}(s)$. The peaks become narrower and larger as $N$ becomes larger. (e) The critical point $s_{c}(N)$ extracted from the peak of the susceptibility plotted against system size, for the DW densities $n_{\text{DW}} = 1/2$ (blue circle) and $n_{\text{DW}} = 1/4$ (red diamonds). We fit the data the the power law $s_{c}(N) \propto N^{-\alpha}$ (solid line) and to the polynomial $s_{c}(N) = aN^{-2} + bN^{-3} + cN^{-4}$ (dashed lines) which are the sub-leading corrections to $N^{-2}$.}
    \label{fig:mps_1}
\end{figure*}

\subsection{Matrix Product States}

A matrix product state (MPS) is an ansatz for the vector state of a many-body system \cite{Vidal2003,Verstraete2004,Schollwoeck2011} .
For a chain of $N$ finite dimensional subsystems (of dimension $d$), it corresponds to states of the form
\beq
    \ket{\Psi} = \sum_{i_{1}, \dots, i_{N}}^{d} \Tr\big( A_{1}^{i_{1}}A_{2}^{i_{2}} \dots A_{N}^{i_{N}} \big) \ket{i_{1} \, i_{2}\, \dots \, i_{N}}
    \label{MPS}
\eeq
where $i_{k}$ labels the states of the physical basis for the $k$-th subsystem and each $A_{k}$ is a rank-3 tensor with dimensions $d\times D\times D$,
with $D$ the so-called bond dimension. Thus the MPS is described by $O(NdD^{2})$ parameters.
Notice that by increasing $D$, any arbitrary state can be exactly written in the form \eqref{MPS}, although this may require up to $D = d^{\lfloor N/2 \rfloor}$.

The bond dimension $D$ limits the entanglement within the state. More precisely, in a MPS with bond dimension $D$, the entanglement entropy for a subchain $L$ (defined as $S_{E} = - \Tr \rho_{L} \log \rho_{L}$ where $\rho_{L} = \Tr_{N \setminus L }\ket{\Psi}\bra{\Psi}$ is the subchain reduced density matrix) is upper-bounded by $S_{E} \leq 2\log D$, independent of the subchain length. 
This implies that the MPS satisfies an entanglement area law, which is intimately related to their success at approximating relevant physical states \cite{Schuch2008}.
In particular, ground states of local gapped Hamiltonians, which in one spatial dimension are known to satisfy an area law \cite{Hastings2007},
but also of critical models, can be efficiently approximated by MPS \cite{Verstraete2006,Hastings2007}.
Furthermore, MPS constitute the basis of efficient numerical methods,
 including the celebrated density matrix renormalization group (DMRG) algorithm \cite{White1992,Schollwoeck2005} which we use to approximate the ground state of $\mathbb{H}_{s}$.

The DMRG, originally formulated in \cite{White1992}, can be understood as a variational minimization of energy over the set of MPS. By writing the operator \er{Hs} as a matrix product operator (MPO)~\cite{McCulloch2007,Pirvu2010a}, one can perform a local optimization on a single tensor within the MPS to minimize the energy. We iterate through each tensor, applying local updates until convergence. This variational MPS search (vMPS) is well detailed in many reviews (e.g. \cite{Verstraete2008, Schollwoeck2011}). For completeness, we give a brief description in the appendix.

When applying the vMPS to study the LD statistics of the XOR-FA model, we use open boundary conditions (OBC) which formally corresponds to setting $n_{0} = n_{N+1} = 0$, 
as this allows for the most efficient MPS calculations, with computational cost $O(D^{3})$. 
In our problem, the number of DWs defines a global conserved quantity, and we need to find the ground state in a particular sector.
Although it is possible to encode this symmetry in the tensors~\cite{PerezGarcia2008sym,Singh2010,PerezGarcia2010njp},
as other local constraints have \cite{Chepiga2019}, we
opt here for a simpler approach.
Namely, we add an energy penalty to the Hamiltonian to favour the desired sector.
Specifically, the penalty is $\lambda(\mathbb{N}_{\rm DW} - N_{\rm DW})^{2}$ where $\lambda > 0$ is some Lagrange multiplier and $\mathbb{N}_{\rm DW} = \sum_{i = 0}^{N} n_{i}(1-n_{i+1}) + (1-n_{i})n_{i+1}$ is the operator which counts the number of DWs.

\begin{figure*}[th]
    \centering
    \includegraphics[width=\linewidth]{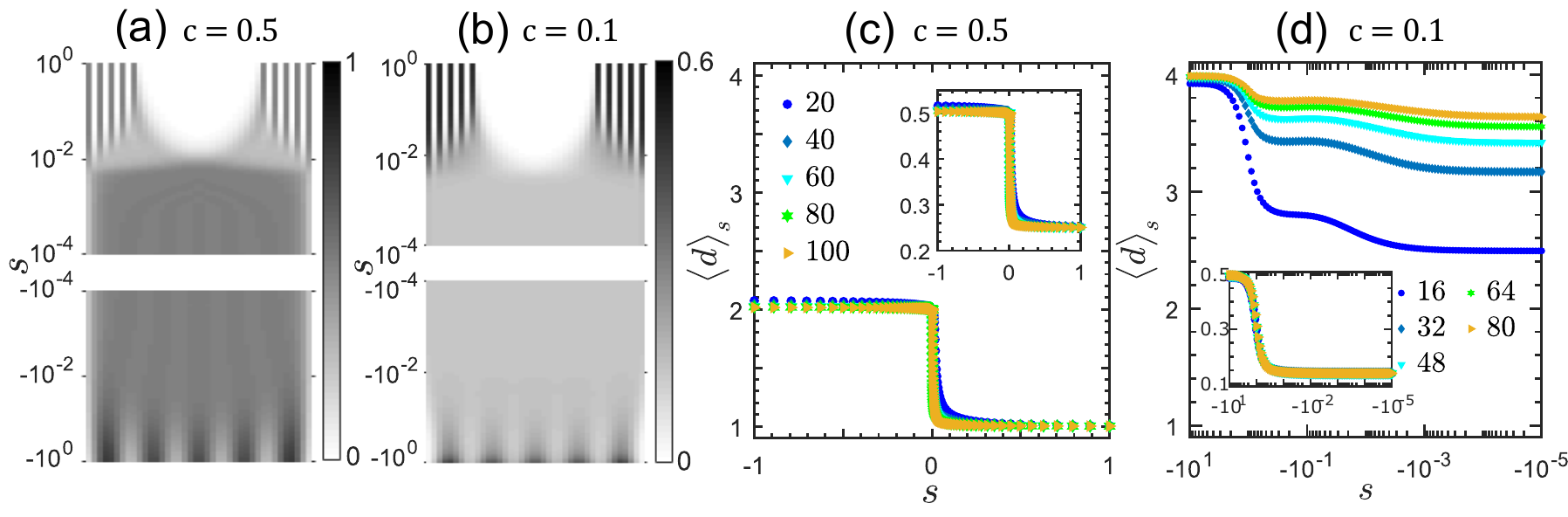}
    \caption{\textbf{Spatial structure of the active and inactive phases.} (a) The average occupation of each site for a system with $c = 0.5$, $N = 40$ and $N_{\rm DW} = 10$. For $s<0$ there is a clear localisation of domains. Each domain becomes (on average) equally sized and hence the DWs are equally spread. For $s>0$, we see the DWs gather at the edge(s). (b) The same as (a) but for $c = 0.1$. (c) The average distance between neighbouring DWs for a systems with $c = 0.5$ and $n_{\rm DW} = 1 / 2$. This shows there is a maximal separation between DWs in the active phase, and only one site separating DWs for the inactive phase. Inset: The average excitation density as a function of $s$. (d) The same as (c) but with $c = 0.1$ and $n_{\rm DW} = 1 / 4$. Here we observe multiple plateaus in the growth. Inset: The average excitation density as a function of s.}
    \label{fig:mps_2}
\end{figure*}

\subsection{Results}

As we now show, the MPS ansatz combined with the variational search proves to be very efficient for studying the LDs of the XOR-FA model, just like for other KCMs and exclusion processes \cite{Banuls2019,Helms2019,Helms2020}. In this way we are able to achieve results for system sizes superior to traditional methods such as exact diagonalisation or importance sampling.

\subsubsection{First-order phase transition in the SCGF} 

A key property of other KCMs such as the FA or the East model is the presence of a first-order dynamical phase transition \cite{Garrahan2007} in the thermodynamical limit $N \rightarrow \infty$, manifested as a singularity in the SCGF $\theta(s)$ at $s = 0$. Consequentially there are two dynamical phases - the active phase for $s < 0$ and the inactive phase for $s > 0$. We look for evidence for this transition in the XOR-FA model.

Figures~\ref{fig:mps_1}(a,b) show the SCGFs obtained numerically for system sizes $N \in [20, 100]$, in linear and log-log scales, respectively. The upper row of Fig.~\ref{fig:mps_1} is for $c=0.5$ while the lower row corresponds to $c=0.1$. For finite size, near enough $s=0$ the SCGF should obey the linear response (LR), $\theta(s) \sim - s k_{\rm eq}$, where $k_{\rm eq}$ is the average activity per unit time in the equilibrium state. For the FA and East models, the equilibrium activity is straightforward to calculate exactly (see e.g.~\cite{Bodineau2012}). For the XOR-FA, it is more difficult due to the conservation of DWs. In the Appendix we give a way to compute it to a good approximation. As we see from Fig.~\ref{fig:mps_1}(b), the SCGF we obtain numerically does obey LR close to $s=0$. 

Still for $s \gtrsim 0$, beyond the LR regime the SCGF changes behaviour, notably stops scaling with system size, see Figs.~\ref{fig:mps_1}(a,b). This change in behaviour becomes even more apparent when one considers the activity per unit time as a function of $s$, $k(s) = - \theta'(s) / N$, Fig. \ref{fig:mps_1}(c). We see a sudden drop close to $s=0$ that becomes more pronounced with system size, a hallmark of a first-order phase transition. From the point of the numerics, this occurs when 
where there two smallest energy levels of \er{Hs} cross. 

The transition point $s_c$ can be estimated from the peak of the susceptibility $\chi(s) = \theta''(s)$, see Fig.~\ref{fig:mps_1}(d). The peaks become higher and sharper with system size. 
From the numerics we can make a finite size scaling analysis of the critical point. We find that 
$s_c(N)$ seems to obey $s_{c}(N) \propto N^{-\alpha}$ as shown in Fig.~\ref{fig:mps_1}(e). For $n_{\rm DW} = 1/2$, we find that the best fit exponent is $\alpha \approx 2.123$ for both $c=0.5$ and $c=0.1$. Furthermore, for $n_{\rm DW} = 1/4$ we find that $\alpha \approx 2.056$ and $\alpha \approx 2.188$ for $c=0.5$ and $c=0.1$ respectively. In each case $\alpha$ is close to the value $2$ expected from a diffusive behaviour of the gap in the spectrum of \er{Hs}. It could be that these are subleading corrections to $N^{-2}$, see Fig.~\ref{fig:mps_1}(e).

\subsubsection{Spatial structure of active and inactive phases}

The singularity in the SCGF represents a phase transition in the trajectories of the dynamics at the level of fluctuations: the behaviour at $s<0$ corresponding to dynamics with activity that is larger than the typical (equilibrium) one is fundamentally different from that at $s>0$ corresponding to dynamics which is less active than typical. This difference is also manifested in the configurations that are predominantly visited by the different trajectories. That is, active and inactive dynamical phases are associated with very different spatial structures. 

We mean the following. At $s=0$ there is no tilting in the ensemble of trajectories which is the one given by the original dynamics. It corresponds to typical behaviour. Dynamics is ergodic over configuration space, and over long-times the distribution of configurations that are visited - for some fixed value of the number of DWs - is given by the equilibrium probability \er{SS}. The state \er{SS} is the leading right eigenstate of generator \er{Ws} at $s=0$. 

At $s \neq 0$ the probability of a trajectory is reweighted by the exponential of its activity, cf.~\er{Zs}. How often configurations are visited in such reweighted ensembles depends on $s$, and for long-times the distribution over configurations is given by the leading eigenstate of \er{Ws}, 
or equivalently \er{Hs} for the detailed balance-obeying XOR-FA. We have access to these states, $\ket{\psi_{s}}$, from our MPS numerics.

The easiest way to characterise the spatial structure of the characteristic configurations associated with dynamics tilted by $s$ is to study the average local occupations $\braket{n_{i}}_s = \braket{\psi_{s} | n_{i} | \psi_{s}}$
\footnote{This only works because here we have OBC. For PBC, one would have to consider the spatial correlations to learn about the structure of the state.}.
In Figs.~\ref{fig:mps_2}(a,b) we show these local densities as a function of $s$ for two values of $c$ and $n_{\rm DW} = 1 / 4$. For $s$ large and negative we see that that domains becomes maximally sized, that is, DWs become maximally spaced apart, maximising the activity, as expected.  In contrast, for $s$ large and positive DWs become localised at either edge of the system. When DWs become minimally separated and clustered together, only the sites next to the last DW are allowed to move and activity becomes subextensive and thus minimal.

We can further quantify the average distance between neighbouring DWs by considering the operator 
\begin{align}
{\mathbb D}_{d} = \sum_{i=1}^{N+1} &n_{i-1}(1-n_{i})(1-n_{i+1}) \dots (1-n_{i+d-1})n_{i+d} \nonumber \\ 
 + &(1-n_{i-1})n_{i}n_{i+1} \dots n_{i+d-1}(1-n_{i+d}) \ ,
\end{align}
which measures the likelihood of observing two neighbouring DWs at distance $d$ apart. The average distance between neighbouring DWs is then given by
\beq
\braket{d}_s = (N_{\rm DW} - 1)^{-1} \sum_{d} \braket{\psi_{s} | {\mathbb D}_{d} | \psi_{s}}
\eeq
(as we have $N_{\rm DW} - 1 $ pairs of neighbouring DWs). 
In Fig.~\ref{fig:mps_2}(c) we show $\braket{d}_s$ as a function of $s$ for $c=0.5$ and  
$n_{\rm DW} = 1 / 2$. It is evident from the plot that the dynamical transition at $s_c \approx 0$ is also one where there is a singular change in the distance between DWs in the corresponding characteristic configurations, from maximal distance between DWs at $s$ negative, to minimal at $s$ positive.

Figure~\ref{fig:mps_2}(d) shows the same for $c=0.1$ and $n_{\rm DW} = 1 / 4$. We see that away from the SEP limit of the XOR-FA, there is even richer spatial structure due to the energetic cost associated with domains. At small $s<0$ there is an initial plateau in the growth of the average distance between DWs. This is an extension of the equilibrium behaviour, where domains are randomly sized according to the behaviour described in Sec. III. As we move further into the active phase, we observe another plateau. At this point, the excitation density $\braket{n}_{s} = N^{-1}\sum_{i}\braket{n_{i}}_{s}$ (as shown in the inset) has not varied much from the equilibrium value. This leads us to believe that this change in structure is due to the excited domains spreading apart and becoming localised as shown in Fig.~\ref{fig:mps_2}(b). This maximises the activity without having to grow the excited domains as is energetically favourable for $c<1/2$. Following this plateau, there is a steady growth to the maximum $\braket{d}_{s}$ corresponding to the growth of the one domains, such that every DW is on average equally spaced. This is further supported by the fact that following this plateau, the excitation density rapidly grows.

\begin{figure}[t]
    \centering
    \includegraphics[width=\linewidth]{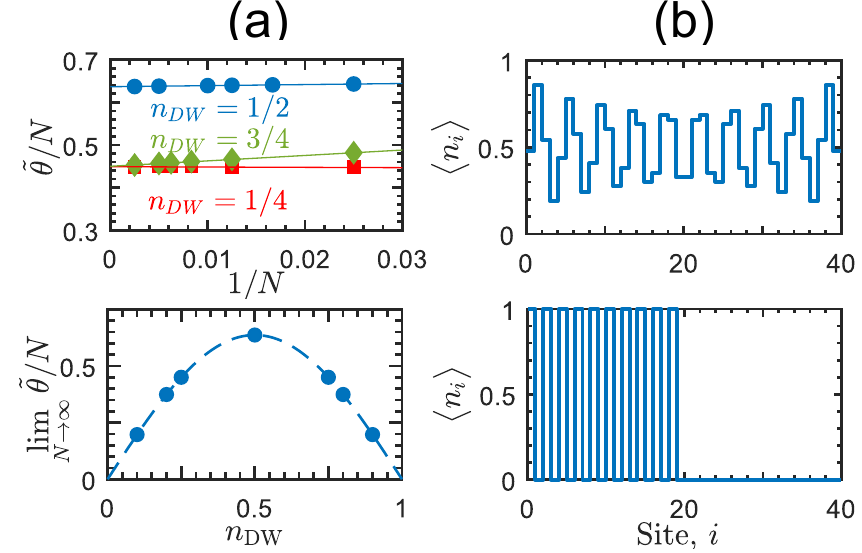}
    \caption{\textbf{Maximally and minimally active limits.} (a) Top: The re-scaled SCGF $\Tilde{\theta}/N = \lim_{s\to-\infty}e^{s}\theta(s) / \sqrt{c(1-c)}$ for various number of DWs. We fit the data with the curves $\Tilde{\theta}/N = a + bN^{-1}$. Bottom: $\Tilde{\theta}/N$ in the limit $N\to\infty$ obtained by extrapolating the fitted curves in top figure, plotted against the density of domain walls $n_{\rm DW}$. We fit the data using $\lim_{N\to\infty}\Tilde{\theta}/N = \frac{\pi}{2}\sin(\pi n_{\rm DW})$. (b) The average occupation at each site for systems a system with $N = 40$ and $N_{\rm DW} = 20$ in the limits $s\to-\infty$ (top) and $s\to\infty$ (bottom). For the latter, we show only one of the degenerate ground states. In this case, DWs localise at the left.}
    \label{fig:mps_3}
\end{figure}

\subsubsection{Maximally and minimally active limits} 

While we cannot calculate the SCGF analytically in general, there are limits where the calculation becomes tractable (apart from the obvious case of $\theta(0)=0$). These are the $s \to \pm\infty$ limits corresponding to the tilting of the dynamics that maximises and minimises the activity. For the former we can easily calculate the ground state of \er{Hs} via vMPS to obtain the rescaled SCGF $\Tilde{\theta} / N = \lim_{s\rightarrow-\infty}e^{s}\theta(s)/(N\sqrt{c(1-c)})$
using only a small bond dimension of $O(10)$. The numerical data is shown in Fig.~\ref{fig:mps_3}(a) 
for various filling fractions. Note that for $s \to -\infty$ the dependence on $c$ is irrelevant, and can be scaled out as in our definition of $\Tilde{\theta}$.

We can extrapolate from the numerical results for finite size to obtain an estimate of $\lim_{N\rightarrow\infty} \, \Tilde{\theta} / N$ as a function of the density of domain walls $n_{\rm DW}$. This large-size limit of the SCGF can be obtained analytically. For $s \to -\infty$, after scaling out the $e^{-s}$ and $\sqrt{c(1-c)}$ factors, the Hamiltonian \er{HDW} becomes that of the XX model. Being non-interacting, the ground state can be obtained exactly by standard means \cite{Pasquale2008}, to give:
$\lim_{N\rightarrow\infty} \, \Tilde{\theta} / N  = \frac{\pi}{2} \sin(\pi \, n_{\rm DW})$. Figure ~\ref{fig:mps_3}(a) shows the agreement between the numerical extrapolation and the exact result. The structure of the state in the maximally active limit for a system with $N = 40$ sites and $N_{\rm DW} = 20$ domain walls is shown in Fig. ~\ref{fig:mps_3}(b). In the limit $N\to\infty$ we would expect that $\lim_{s\to-\infty}\braket{n}_{s} = 1/2,$ and $\lim_{s\to-\infty}\braket{d}_{s} = {n_{\rm DW}}^{-1}$ which are both in excellent agreement with the numerical vMPS data.

For the minimally active limit $s\to\infty$, the Hamiltonian given in Eq.~(\ref{Hs}) becomes a diagonal one. Each configuration is an eigenstate and one can easily show that the configuration with the least energy is the one where all the DWs are clustered at the edge of the system (which is doubly degenerate). The SCGF at this limit is given by $\theta(s\to\infty) = -c$ and the exact structure for $N=40$ and $N_{\rm DW}=20$ is given in Fig.~\ref{fig:mps_3}(b) for just one of the degenerate states (in practice the vMPS prefers to converge onto just one to keep the bond dimension minimal). Additionally, the excitation density and the distance between DWs are given by $\lim_{s\to\infty} \braket{n}_{s}, \, \braket{d}_{s} = n_{\rm DW} / 2, \, 1$ respectively.

\section{Comparison between the FA, XOR-FA and PXP models}

As discussed above, the kinetic constraint of the XOR-FA model is stronger than that of the FA model (which is a binary OR operation on the spins neighbouring the one attempting to flip), but weaker than that of the PXP model (which is a binary AND operation on the neighbouring spins). This has significant consequences on the dynamics. 

\begin{figure}[t]
    \centering
    \includegraphics[width=\linewidth]{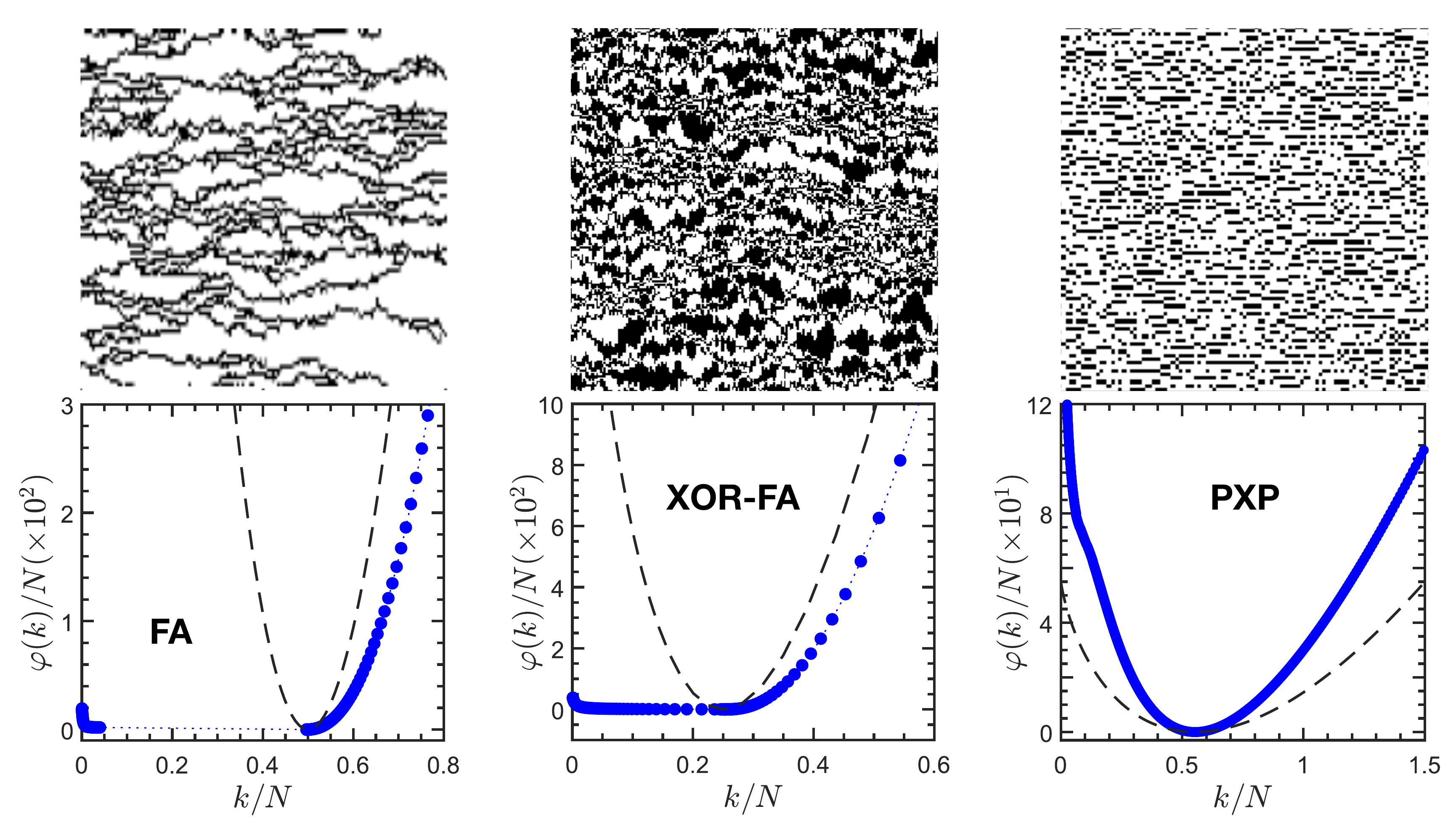}
    \caption{\textbf{Comparison between FA, XOR-FA and PXP models.}
    The top panels show sample equilibrium trajectories for the FA model (left, taken from \cite{Garrahan2002}, $c\approx0.27$, $N=100$), the XOR-FA (middle, $1/4$ filling, $c=0.5$, $N=200$), and the PXP model (right, no pairs of contiguous up spins, $c=0.5$, $N=100$). 
    The bottom panels show the corresponding LD rate functions for the activity in the three models ($c=0.5$ for each) obtained via numerical MPS. The data for the FA model is from \cite{Banuls2019} ($N=200$), and that for the PXP is from \cite{Chepiga2019} ($N=400$). Both the FA and XOR-FA ($1/2$ filling, $N= 100$) rate functions show first-order transitions close to the typical dynamics (flatness of the bottom of the curves), while the PXP has a continuous transition for highly atypical low values of the activity (kink at left of curve, see \cite{Chepiga2019}). The dashed curves are Poisson rate functions for comparison. 
    }
    \label{fig:comparison}
\end{figure}

In the case the FA model \cite{Ritort2003}, configuration space is all connected by the dynamics, except for the configuration with $n_i=0$ for all $i$. This means that in practice dynamics is irreducible and there is one giant ergodic component (as the probability of starting from the unique all-zero site is suppressed exponentially  with size). Despite the relative weakness of the constraint, the dynamics of the FA is strongly fluctuating \cite{Garrahan2002}. Figure \ref{fig:comparison} (top left) shows an example trajectory in equilibrium, displaying the characteristic ``space-time bubbles'' responsible for dynamic heterogeneity \cite{Garrahan2003,Chandler2010}. This preponderance of fluctuations is manifest in the form of the LD rate function for the dynamical activity, see Fig.~\ref{fig:comparison} (bottom left), corresponding to a (dynamical) first-order transition \cite{Garrahan2007}.

On the other extreme of this comparison is the PXP model
\cite{Fendley2004,Lesanovsky2011,Bernien2017,Turner2018}. 
In the stochastic terminology this corresponds to the 2-spin facilitated FA model in one-dimension \cite{Ritort2003}. 
Here the constraint is such that configuration space is broken into exponentially many dynamically disconnected components. Specifically, pairs of sites with up spins, $n_i = n_{i+1} = 1$ cannot be flipped and are conserved by the dynamics. This means that dynamics is reducible as configuration space is partitioned into an exponential in size number of irreducible components, classified by local conservation laws (i.e., the location of the unmovable contiguous clusters of up spins). The largest ergodic component is that where no two up spins are adjacent. But despite the strength of the constraint, the resultant dynamics is much less interesting than for the FA or the XOR-FA models, see for example the sample trajectory of Fig.~\ref{fig:comparison} (top right). Correspondingly, a detailed quantification of the statistics of the dynamics shows no significant fluctuations, see the LD rate function of Fig.~\ref{fig:comparison} (bottom right).

The middle panels of Fig.~\ref{fig:comparison} show the XOR-FA for comparison. Given that its constraint is somewhat in between that of the FA and PXP model, we see that trajectories display less pronounced ``bubbles'' than the FA but are more fluctuating that the PXP.
Specifically, the constraint does break configuration space, but the number of disconnected ergodic components is only linear in the system size.
These components are classified by the globally conserved number of domain walls.
This allows for rich dynamics within the components, leading to singular LD behaviour as demonstrated in this paper; see LD rate function Fig.~\ref{fig:comparison} (middle bottom).  

The key observation is that both FA and XOR-FA models have trajectory transitions which manifest in fluctuating dynamics, while the PXP does not. In both the FA and XOR-FA models there are configurations which are completely void of dynamics, the all-zero state in the FA, and the state with maximal number of domain walls in the XOR-FA. But while these configurations are disconnected dynamically, many other configurations have finite regions that resemble them locally. Such spatial rare regions can only be relaxed from the outside, and thus give rise to the dynamical bubbles of the trajectories, see Fig.~\ref{fig:comparison}, and are at the source of the large fluctuations in the dynamics. 
In contrast, the PXP constraint makes inactive configurations local rather than global, and they can either be relaxed locally or not. In the PXP there are no space-time bubbles and no LD transition close to $s=0$ (close at the edge of typical dynamics, see discussion of previous sections).

\section{Conclusions}

We have studied a new classical stochastic KCM, the one-dimensional XOR-FA model, which could be experimentally realized with a laser-driven dissipative Rydberg lattice gas under facilitation (anti-blockade) conditions \cite{lesanovsky2013, Lesanovsky2014, Hoening2014}. The kinetic constraint in this model is stronger than that of the standard FA model, as spins can flip if only one nearest neighbour is in the excited state, in contrast to the FA where spins can also flip if more than one neighbour is excited. It is also less constrained than the PXP, or 2-spin facilitated FA, model which requires both neighbours to be simultaneously in the same state. As such it shares features with both these models. The constraint imposes a conservation law, that of the total number of domain walls, breaking configuration space into a number of disconnected components that scales with system size. This contrasts to the FA model where all but one configuration are dynamically connected, and is closer to the PXP where configuration space is also disconnected. The distinction with the PXP is that in the XOR-FA the conserved quantity is global, while in the PXP there are many local conserved quantities and configuration space is broken into exponentially many disconnected components. This less severe disconnection of state space makes the dynamics within connected components in the XOR-FA still interesting as there is scope for large dynamical fluctuations (in contrast to the PXP). An interesting question would be to study the analogous problem in constrained lattice gases, considering variants of the Kob-Andersen (KA) model \cite{Kob1993} or the triangular lattice gas (TLG) \cite{Jackle1994} but where constraints are ``selective'' (as in Ref.~\cite{Sellitto2019}). The KA and the TLG are models where particles can hop as long as a minimum number of neighbours are unoccupied (cf.\ the FA model where at least one neighbour has to be up for the spin to flip). A ``selective'' \cite{Sellitto2019} version of such models where the number of required unoccupied neighbours is fixed would be slightly more constrained, just like the XOR-FA is slightly more constrained than the FA model. Studying such models in the manner of the current paper would require however to extend the tensor network methods to dimensions higher than one.

The XOR-FA is a ``thermal'' model, in the sense that it obeys detailed balance with respect to the Boltzmann distribution of a non-interacting binary gas, where each excited spin costs a unit of energy, and subject to the conservation of the number of DWs. In analogy to other KCMs like the FA and East models \cite{Ritort2003}, the XOR-FA has a trivial (up to the conservation law) thermodynamics, but complex relaxation dynamics due to the constraint. The conservation law allows the XOR-FA to be represented in terms of the dynamics of its DWs. This establishes a connection to exclusion processes. At infinite temperature the XOR-FA can be mapped via a duality transformation to the SEP \cite{Ostmann2019,Suzuki1971,Borla2020}, while at finite temperature this mapping leads to a dynamics of hopping DWs with long-range interactions (in contrast to the original spin representation which is always local).

Like in other KCMs \cite{Garrahan2007,Garrahan2009,Bodineau2012,Garrahan2018}, we have shown here that the XOR-FA has a trajectory level phase transition between active and inactive dynamical phases. We have proved this to high accuracy via numerical MPS. This adds to recent work \cite{Banuls2019,Helms2019,Helms2020} demonstrating the effectiveness of numerical tensor network methods for studying the full counting statistics of stochastic systems. One of the many advantages of this approach is the direct availability of spatial structural information in the various dynamical phases. For the XOR-FA we find spatial structure as expected from its connection to the SEP \cite{Jack2015,Lecomte2012}: a change from repulsion of DWs in the active phase, maximising the possibility of motion, through structureless configurations in the typical equilibrium dynamics, to DW clustering in the inactive phase. Away from the strict SEP limit, these structures remain overall, with further richness due to energetics. 

Here we have studied the XOR-FA in one-dimension. It is easy to generalise the model to higher dimensions, once again motivated for example by the problem of atoms interacting strongly in Rydberg states. For example, in the blockade regime, when going from one dimension to a, say, two dimensional square lattice, the PXP model becomes the hard square model \cite{Ji2011}. Similarly, in the anti-blockade regime of Rydbergs, the XOR-FA would generalise to a model that is less constrained than that of hard squares, but more constrained than the 2-spin facilitated FA model in two dimensions \cite{Ritort2003}. From the results here it is safe to speculate that such higher dimensional generalisations of the XOR-FA will also display very rich dynamics. Additionally, the structure of the ground state observed at $s>0$ is reminiscent of the localized ground states found in quantum KCMs \cite{Pancotti2020}, which has dramatic consequences for the quantum dynamics of the model. It may be interesting to investigate whether an analogous localization can  be characterized in the ground state of the XOR-FA model in the inactive regime.

\acknowledgments
We thank Oriane Blondel, Martin Evans and Sarang Gopalakrishnan for useful comments on the manuscript. We acknowledge financial support from EPSRC Grant no.\ EP/R04421X/1 and the Leverhulme Trust Grant No. RPG-2018-181.
I.L.\ acknowledges support from the DFG through SPP 1929 (GiRyd) and by the Wissenschaftler-R\"uckkehrprogramm GSO/CZS of the Carl-Zeiss-Stiftung and the German Scholars Organization e.V.
M.C.B.\ acknowledges support from Deutsche Forschungsgemeinschaft (DFG, German Research Foundation) under Germany's Excellence Strategy -- EXC-2111 -- 390814868.
We are grateful for access to the University of Nottingham Augusta HPC service.

\bibliographystyle{apsrev4-1}

\newpage

\appendix

\section*{APPENDIX}

\subsection*{Variational MPS}
\begin{figure}[t]
    \centering
    \includegraphics[width=\linewidth]{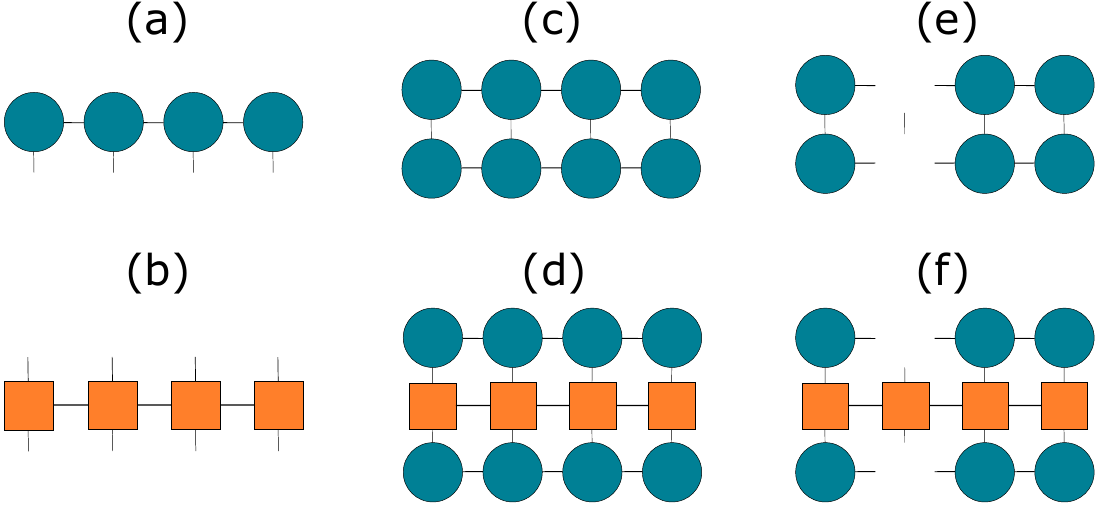}
    \caption{\textbf{Tensor networks in the vMPS.} The diagrammatic representations of tensor networks in the variational MPS search. (a) The state $\ket{\Psi}$ as an MPS. (b) The Hamiltonian $\mathbb{H}_{s}$ as an exact MPO. (c) The inner product $\braket{\Psi | \Psi}$. (d) The expectation value $\braket{\Psi | \mathbb{H}_{s} | \Psi}$. (e) The effective norm $\mathcal{N}_{k}$. (e) The effective Hamiltonian $\mathcal{H}_{k}$.}
    \label{fig:mps_diagrams}
\end{figure}

The vMPS algorithm used in Sec.~IV goes as follows. We have some MPS, $\ket{\Psi_{\rm guess}}$ as defined in \er{MPS}, which is our guess to the true ground state. See Fig.~\ref{fig:mps_diagrams}(a) for the diagrammatic representation, where the shapes represent the local tensors and the legs represent contractions over tensors.  One can then write the Hamiltonian in \er{Hs} as a matrix product operator (MPO) \cite{McCulloch2007,Pirvu2010a}
\begin{align}
    \mathbb{H}_{s} = \sum_{i_{1}, .., i_{N}}\sum_{j_{1}, .., j_{N}}
    & \Tr(M_{1}^{i_{1}j_{1}} M_{2}^{i_{2}j_{2}}\, \dots\, M_{N}^{i_{N}j_{N}}) 
    \label{HsMPO}
    \\
    & \ket{i_{1} \, i_{2}\, \dots \, i_{N}}\bra{j_{1} \, j_{2}\, \dots \, j_{N}} \nonumber
\end{align}
where $M_{k}$ is a rank-4 tensor with dimensions $d \, \times \, d \, \times \, D_{H} \, \times \, D_{H}$. The locality of $\mathbb{H}_{s}$ allows us to exactly represent it in MPO form with only a small bond dimension $D_{H} = 4$, where each tensor is identical. As with the MPS, this can be represented in the diagrammatic form Fig.~\ref{fig:mps_diagrams}(b). The energy of the guess with respect to \er{HsMPO} is then given by
\beq
    E_{\rm guess} = \frac{\braket{\Psi_{\rm guess} | \mathbb{H}_{s} | \Psi_{\rm guess}}}
    {\braket{\Psi_{\rm guess} | \Psi_{\rm guess}}} \geq E_{s}
    \label{Eguess}
\eeq
where $E_{s} = -\theta(s)$ is the true ground state energy. 
The expectation value and inner product can be expressed as tensor network contractions, as illustrated in Figs.~\ref{fig:mps_diagrams}(c, d). 
This allows for an efficient calculation that exploits the MPS structure.

At each step, a single tensor is optimized by minimizing equation \er{Eguess} with respect to $A_{k}$, which gives
\beq
    \mathcal{H}_{k} A_{k} = E_{\rm guess} \, \mathcal{N}_{k} A_{k}.
    \label{gep}
\eeq
where $\mathcal{N}_{k}$ and $\mathcal{H}_{k}$ are the effective norm and effective Hamiltonian computed by contracting over all tensors except for $A_{k}$ within $\braket{\Psi_{\rm guess} | \Psi_{\rm guess}}$ and $\braket{\Psi_{\rm guess} | \mathbb{H}_{s} | \Psi_{\rm guess}}$ respectively. 
Both effective operators can be expressed also as tensor networks, as shown in Figs.~\ref{fig:mps_diagrams}(e, f). If we treat $A_{k}$ as a $(D^{2}d)$-vector and $\mathcal{N}_{k}$, $\mathcal{H}_{k}$ as  $(D^{2}d)$-matrices, then \er{gep} is simply a generalized eigenvalue problem which should be solved using a sparse eigensolver to keep the computational scaling to $O(D^{3})$. The solution to \er{gep} with the smallest $E_{\rm guess}$ is our new choice for $A_{k}$.

We sweep back and forth through each tensor in the MPS, applying local updates in the way detailed above.
Since each local minimization can be solved exactly, the energy can only decrease at each step, and the algorithm is guaranteed to converge. 
However, it may do so to a local minimum. 
As a quality criterion, we require that the (efficiently computable) variance of the Hamiltonian in the guess state
falls below some specified value $\braket{\mathbb{H}_{s}^{2}} - \braket{\mathbb{H}_{s}}^{2} < \epsilon$, where here $\braket{\cdot}$ denotes an expectation with respect to the $\ket{\Psi_{\rm guess}}$. If $\ket{\Psi_{\rm guess}}$ does not satisfy this criterion for a run of the algorithm at some bond dimension $D$, we run it again with an MPS with a higher bond dimension, typically using the previous run as our initial guess.

\subsection*{MPS steady state solutions}
Here we follow the workings of Ref.~\cite{Sidoravicius1998} to present an MPS steady state solution to \er{SS} in the thermodynamic limit, $N\to\infty$, with which we can determine the equilibrium properties of the model.
We consider the XOR-FA model with $N$ sites and PBC, and describe the probability vector $\ket{\nu}$ as an MPS, c.f. \er{MPS}, where of course we have translational invariance and each tensor is identical, $A_{k} = A$ for all $k$. Let us now guess the solution
\begin{equation}
    A^{0} = \begin{bmatrix}
    0 & 0\\
    1-p_{0} & \;p_{0} \;
    \end{bmatrix}, \quad
    A^{1} = \begin{bmatrix}
    \;p_{1}\; & 1 - p_{1}\\
    0 & 0
    \end{bmatrix},
\end{equation}
where $0 < p_{0}, \,  p_{1} < 1$. We first require that our solution annihilates the generator, $\W\ket{\nu} = 0$. It is easy to verify that this is case if we have
\begin{equation}
    \frac{p^{1}}{p^{0}} = \frac{c}{1-c}.
    \label{appendix_relation}
\end{equation}
Additionally, we require that $\ket{\nu}$ is normalised.
The partition function is calculated by taking the inner product with the flat state,
\beq
    Z = \braket{- | \nu} = \Tr(M^{N}),
\eeq
where $M^{N} = A^{0} + A^{1}$. It is easy to show via induction that
\begin{align}
    M^{t} = &(2-p_{0}-p_{1})^{-1} \Bigg(
    \begin{bmatrix}
    1 - p_{0} & 1 - p_{1} \\
    1 - p_{0} & 1 - p_{1}
    \end{bmatrix}
    \nonumber
    \\
    &+ (p_{0} + p_{1} - 1)^{t} 
    \begin{bmatrix}
    1 - p_{1} & -(1 - p_{1}) \\
    -(1 - p_{0}) & 1 - p_{0}
    \end{bmatrix}
    \Bigg) .
    \label{appendix_identity}
\end{align}
It follows that the partition function is already normalised in the infinite limit
\beq
    \lim_{N\to\infty} Z = 1 .
\eeq

The average DW density $\braket{n_{\text{DW}}}$ can be calculated as the DW occupation between any two neighbouring sites in the lattice,
\begin{align}
    \braket{n_{\text{DW}}} &= \frac{1}{Z} \braket{- | \big[n_{i}(1-n_{i+1}) + (1-n_{i})n_{i+1}\big] | \nu}
    \nonumber
    \\
    & = \frac{1}{Z} \Tr\left(\left[ A^{1}A^{0} + A^{0}A^{1} \right] M^{N-2} \right).
\end{align}
Taking the infinite limit, one finds that 
\beq
    \lim_{N\to\infty}\braket{n_{\text{DW}}} =  \frac{2(1-p_{0})(1-p_{1})}{2-p_{0}-p_{1}} .
    \label{appendix_DW}
\eeq
We can now determine the $p_{0}, \, p_{1}$ necessary to have the required DW density by substituting \er{appendix_relation} into \er{appendix_DW} and solving as a quadratic equation.
Thus we have found an MPS steady state solution with bond dimension $2$, which also has the required DW density in the thermodynamic limit. 

To calculate other local observables, we can again simply use the procedure described above.
The average excitation density can be calculated using a local MPO on just one site
\begin{align}
    \lim_{N\to\infty} \braket{n} &= \lim_{N\to\infty} \frac{1}{Z} \braket{- | n_{i} | \nu}
    = \lim_{N\to\infty} \frac{1}{Z} \Tr \left( A^{1}M^{N-1} \right)
    \nonumber
    \\
    &= \frac{1-p_{0}}{2-p_{0}-p_{1}} .
\end{align}
Likewise, the average dynamical activity can be calculated as the escape rate of just a single site, which can be calculated using the three-body operator
\begin{align}
    r_{i} &= c\big[n_{i-1}(1-n_{i})(1-n_{i+1}) + (1-n_{i-1})(1-n_{i})n_{i+1}\big]
    \nonumber
    \\
    &+ (1-c)\big[n_{i-1}n_{i}(1-n_{i+1}) + (1-n_{i-1})n_{i}n_{i+1}\big] .
\end{align}
After a lengthy calculation, we find
\begin{align}
    \lim_{N\to\infty} \braket{k} &= \lim_{N\to\infty} \frac{1}{Z} \braket{- | r_{i} | \nu}
    \nonumber
    \\
    &=  \frac{2(1-p_{0})(1-p_{1})}{2-p_{0}-p_{1}}\big[(1-c)p_{1} + cp_{0}\big] .
\end{align}
We compare our analytical results to numerical data obtained for large, but finite system sizes in Fig.~\ref{fig:relaxation1}. Both results show excellent agreement.

\end{document}